\documentclass[aps,prb,preprint,showpacs,preprintnumbers,footinbib,amsmath,amssymb]{revtex4-1}

\usepackage[dvipdfmx]{graphicx}
\usepackage{bm}

\newcommand{\be}{\begin{equation}}
\newcommand{\ee}{\end{equation}}
\newcommand{\eq}[1]{Eq.~(\ref{#1})}
\newcommand{\fig}[1]{Fig.~\ref{#1}}
\def\bea{\begin{eqnarray}}
\def\eea{\end{eqnarray}}

\def\bra{\langle}
\def\ket{\rangle}
\def\vq{{\bf q}}

\def\vk{{\bf k}}

\def\vr{{\bf r}}
\def\qp{{\bf q}_{\parallel}}

\begin{document}

\title{Dual structure in the charge excitation spectrum of electron-doped cuprates} 

\author{Mat\'{\i}as Bejas$^\dag$, Hiroyuki Yamase$^\ddag$, and Andr\'es Greco$^\dag$}
\affiliation{
{$^\dag$}Facultad de Ciencias Exactas, Ingenier\'{\i}a y Agrimensura and
Instituto de F\'{\i}sica Rosario (UNR-CONICET),
Av. Pellegrini 250, 2000 Rosario, Argentina\\		
{$^\ddag$}National Institute for Materials Science, Tsukuba 305-0047, Japan
}

\date{\today}

\begin{abstract}
Motivated by the recent resonant x-ray scattering (RXS) and resonant inelastic x-ray
scattering (RIXS) experiments for electron-doped cuprates, we study 
the charge excitation spectrum in a layered $t$-$J$ model with the long-range Coulomb interaction. 
We show that the spectrum is not dominated by  a specific type of charge excitations, but 
by different kinds of charge fluctuations, and is characterized by a dual structure in the energy space. 
Low-energy charge excitations 
correspond to various types of bond-charge fluctuations 
driven by the exchange term ($J$-term)  
whereas high-energy charge excitations are
due to usual on-site 
charge fluctuations and correspond to plasmon excitations above 
the particle-hole continuum. 
The interlayer coupling, which is frequently neglected in many theoretical studies, 
is particularly important to the high-energy charge excitations. 
\end{abstract}

\maketitle
\section{introduction}

The origin of the charge order (CO), originally observed in underdoped 
hole-doped cuprates (h-cuprates) \cite{wu11, ghiringhelli12, chang12, achkar12, blackburn13, leboeuf13, blanco-canosa14, comin14, da-silva-neto14, hashimoto14, tabis14, gerber15, peng16, tabis17}, is an active topic
in condensed matter physics. 
The recent x-ray experiments for electron-doped 
cuprates  (e-cuprates) \cite{da-silva-neto15,da-silva-neto16} push the field even further. 
A CO occurs in an intermediate doping region and has 
a modulation vector along the $(0,0)$-$(\pi,0)$ direction. 
These features are very similar to those observed in h-cuprates and 
suggest a common origin. 
However, this may require more careful studies because several  
experiments indicate a large asymmetry between e- and h-cuprates.
For instance, the superconducting critical temperature $T_c$ is much lower in 
e-cuprates than h-cuprates, and the pseudogap is weak or negligible in e-cuprates \cite{armitage10}.
The last fact makes, in principle, e-cuprates simpler for a study 
of charge excitations because the CO may be formed from a homogeneous paramagnetic state and
not from the pseudogap phase as in h-cuprates.

The charge excitation spectrum can be observed by resonant inelastic x-ray scattering (RIXS). 
The energy resolution of RIXS is presently about $50-100$ meV and low-energy excitations are not 
sharply resolved 
compared with the high-energy ones. Given that the CO phenomena are associated with 
low-energy charge excitations, a natural question may arise: 
are the high-energy charge excitations observed by RIXS related with 
the low-energy charge excitations possibly associated with the CO?

This question can be studied for e-cuprates, 
where both the CO \cite{da-silva-neto15,da-silva-neto16} and high-energy charge 
excitations \cite{wslee14,ishii14} are observed in the same system. 
The CO was discussed in terms of resonant x-ray scattering (RXS). 
RXS measures the equal-time correlation function, that is, 
the spectral weight is integrated up to infinity with respect to energy. 
Thus, RXS cannot easily distinguish between a static order and a short-range fluctuating order. 
Refs.~\onlinecite{da-silva-neto15} and \onlinecite{da-silva-neto16} reported a short-range CO 
with a modulation vector $\qp \approx (0.5\pi,0)$; here $\qp$ denotes in-plane momentum. 
In contrast to the case in h-cuprates\cite{allais14,meier14,wang14,atkinson15,yamakawa15,mishra15,bejas12}, 
the CO in e-cuprates has been studied only in a few theories \cite{yamase15b,li17}.  
In Ref.~\onlinecite{yamase15b} 
it was discussed that the observed short-range CO is a $d$-wave bond order and 
its modulation vector is determined by $2k_F$ scattering processes 
that connect the two Fermi momenta on the boundary of the Brillouin zone (BZ). 
The $d$-wave bond order is also proposed in Ref.~\onlinecite{li17} by employing a model 
different from that in Ref.~\onlinecite{yamase15b}.  
On the other hand, high-energy spectrum revealed 
a quasi-linear dispersive mode around $\qp = (0,0)$ with an energy gap 
about $300$ meV (Ref.~\onlinecite{wslee14}). 
The origin of this mode is under debate\cite{greco16,ishii14,ishii17,wslee14}, 
and its relation with the CO phenomena is an open issue. 
A theoretical study of a layered  $t$-$J$-$V$ model \cite{greco16} implies that 
this mode can be plasmons with a finite out-of-plane momentum $q_z$ and originates 
from the usual on-site charge excitations. 

From previous studies\cite{yamase15b,greco16,li17}, it seems that 
the low-energy CO physics and the high-energy charge excitations are of different nature in e-cuprates. 
However, more work would be necessary before reaching such a conclusion. First, the CO was studied 
in the two-dimensional (2D) $t$-$J$ model \cite{bejas14,yamase15b}, whereas the high-energy mode was analyzed 
in the layered $t$-$J$-$V$ model \cite{greco16}, with $V$ being the long range Coulomb interaction. 
Thus, it is not clear whether the obtained insight about the CO in the 2D $t$-$J$ model 
remains valid also in a more realistic model such as the layered $t$-$J$-$V$ model. 
Second, the CO was studied in Ref.~\onlinecite{yamase15b} by focusing on a $d$-wave bond-order instability, 
which is a certain type of charge fluctuations. 
Possible contributions from other CO tendencies \cite{bejas14} were neglected. 
In particular, even if other CO tendencies are not leading, 
charge excitations associated with them could appear in a finite energy region. 
This possibility has never been studied. 
Third and most importantly, a  global understanding of charge 
excitations has not been obtained in cuprates. If that is established in e-cuprates, 
it will certainly be an important step toward the understanding of charge excitations in h-cuprates. 
Therefore it is a challenge to have a theory that may describe both low- and high-energy 
charge excitations on an equal footing and elucidate the charge excitation spectrum 
in a wide range of energy and momentum. 

In this paper we study the momentum and energy resolved charge excitation 
spectrum of a layered  $t$-$J$-$V$ model in large-$N$ expansion. 
The large-$N$ analysis provides a nonperturbative scheme where 
all possible charge channels allowed by symmetry can be handled  
on an equal footing, that is, a certain specific charge 
channel is not considered favorably by hand. 
This feature is particularly important to the present work because 
it is not clear what kinds of charge excitations are actually detected in x-rays experiments. 

We find that various types of charge fluctuations contribute to the charge excitation spectrum, which is 
characterized by a dual structure in the energy space. 
The spectrum in the low-energy region is triggered by the charge sector of the exchange
interaction $J$ and originates from various bond-charge fluctuations, which 
may lead to the observed CO features \cite{da-silva-neto15,da-silva-neto16}. 
On the other hand, the spectrum in the high-energy region 
is dominated by usual on-site 
charge fluctuations and is distinct from the spectrum of the low-energy 
bond-charge fluctuations. In particular, it is dominated by plasmons with a finite 
momentum transfer along the $z$ direction. 

The present paper is organized as follows. 
We first summarize our theoretical scheme in Sec.~II, and then present results for the 
charge excitations in a realistic layered $t$-$J$-$V$ model in Sec.~III, 
including a comparison with the standard 2D $t$-$J$-$V$ model. 
Our obtained results are discussed in Sec.~IV in light of experimental data and 
conclusions are given in Sec.~V. 
Appendices discuss the  $q_z$ dependence of charge excitations and 
an alternative definition of bond-charge susceptibility. 

\section{Theoretical scheme}
Since high-$T_c$ cuprate superconductors are doped Mott insulators, a minimal model 
is the 2D $t$-$J$ model on a square lattice. However, as we discuss in the present paper, 
this minimal model is not sufficient to describe all features observed in the charge excitation spectrum  
of e-cuprates. Two additional effects, which are frequently neglected in many theoretical studies, 
are indispensable: weak interlayer coupling along the $z$ direction and the long-range Coulomb interaction. 
Hence we study the $t$-$J$ model on a square lattice by including interlayer hopping and 
the long-range Coulomb interaction: 
\begin{equation}
H = -\sum_{i, j,\sigma} t_{i j}\tilde{c}^\dag_{i\sigma}\tilde{c}_{j\sigma} + 
\sum_{\langle i,j \rangle} J_{ij} \left( \vec{S}_i \cdot \vec{S}_j - \frac{1}{4} n_i n_j \right)
+ \sum_{\langle i,j \rangle} V_{ij} n_i n_j \, 
\label{tJV}  
\end{equation}
where the sites $i$ and $j$ run over a three-dimensional lattice. 
The hopping $t_{i j}$ takes a value $t$ $(t')$ between the first (second) nearest-neighbors 
sites on the square lattice. The hopping integral between layers is scaled by $t_z$ 
and the electronic dispersion is specified later [see \eq{Ek}]. 
$\langle i,j \rangle$ denotes a nearest-neighbor pair of sites, and 
we consider the exchange interaction only inside the plane, namely $J_{i j}=J$,  
because the exchange term between the planes ($J_\perp$) is much smaller than $J$
(Ref.~\onlinecite{thio88}).
$V_{ij}$ is the long-range Coulomb interaction on the lattice and is given in momentum space by 
\be
V(\vq)=\frac{V_c}{A(q_x,q_y) - \cos q_z} \,,
\label{LRC}
\ee
where $V_c= e^2 d(2 \epsilon_{\perp} a^2)^{-1}$ and 
\be
A(q_x,q_y)=\frac{\tilde{\epsilon}}{(a/d)^2} (2 - \cos q_x - \cos q_y)+1 \,.
\ee
These expressions are easily obtained by solving the Poisson's equation on the square lattice \cite{becca96}.  
Here $\tilde{\epsilon}=\epsilon_\parallel/\epsilon_\perp$,  
and $\epsilon_\parallel$ and $\epsilon_\perp$ are the 
dielectric constants parallel and perpendicular to the planes, respectively; 
$e$ is the electric charge of electrons;  
$a$ is the lattice spacing in the planes and the in-plane momentum $\qp=(q_x,q_y)$ is measured 
in units of $a^{-1}$; similarly $d$ is the distance between the planes and the 
out-of-plane momentum $q_z$ is measured in units of $d^{-1}$.  
$\tilde{c}^\dag_{i\sigma}$ ($\tilde{c}_{i\sigma}$) is 
the creation (annihilation) operator of electrons 
with spin $\sigma$ ($\sigma = \downarrow$,$\uparrow$) 
in the Fock space without double occupancy. 
$n_i=\sum_{\sigma} \tilde{c}^\dag_{i\sigma}\tilde{c}_{i\sigma}$ 
is the electron density operator and $\vec{S}_i$ is the spin operator.

Since the Hamiltonian (\ref{tJV}) is defined in the Fock space without double occupancy, 
its analysis is not straightforward. Here we employ a large-$N$ technique in a path 
integral representation of the Hubbard $X$ operators \cite{foussats04}. 
This formalism allows us to study all possible charge instabilities 
already at leading order  \cite{bejas12}. Furthermore their dynamics turns out to capture essential features of 
charge dynamics observed in electron-doped cuprates \cite{bejas14,yamase15b,greco16,greco17}.  
Details of the formalism are summarized in Ref.~\onlinecite{bejas12} in the case of the 
2D $t$-$J$ model. It is easily extended to the present layered model as 
actually done in Ref.~\onlinecite{greco16}. Hence we present only the essential part 
of our formalism here. 

In the path integral formalism, our model (\ref{tJV}) can be written in terms of a six-component 
bosonic field 
\be
\delta X^a_{i}=(\delta R_{i}, \delta \lambda_{i}, r^x_{i}, r^y_{i}, A^x_{i}, A^y_{i}), 
\label{bosons}
\ee
fermionic fields, and interactions between bosonic and fermionic fields \cite{bejas12}. 
$\delta R_i$ describes on-site charge fluctuations because it comes from 
$X^{00}_i=N\frac{\delta}{2} (1+\delta R_i)$ where $X^{00}_i$ is the Hubbard operator 
associated with the number of holes at a site $i$; 
$\delta$ is the doped carrier density; the factor $N$ comes from 
the sum over the $N$ fermionic channels after the extension of the spin index 
$\sigma$ from $2$ to $N$. 
$\delta \lambda_i$ describes fluctuations of the 
Lagrangian multiplier introduced to impose the non-double occupancy at any site. 
$r^x_i$ and $r^y_i$ ($A^x_i$ and $A^y_i$) are fluctuations of the real (imaginary) part of 
a bond field along the $x$ and $y$ direction, respectively. 
This bond field is a Hubbard-Stratonovich field to decouple the exchange interaction 
in the model (\ref{tJV}) and is parametrized as 
$\Delta_i^{x(y)}=\Delta(1+r_i^{x(y)}+i A_i^{x(y)})$, where 
$\Delta$ is the mean-field value of the bond field, 
which is determined self-consistently \cite{bejas12}, and is proportional to $J$. 
Thus the bond-charge fluctuations naturally appear in the present model. 
After Fourier transformation, the quadratic term of $\delta X^{a}$ defines a $6 \times 6$ 
bare bosonic propagator $D_{ab}^{(0)}(\vq, {\rm i}\omega_n)$, which is given by 
\begin{widetext}
\begin{equation} \label{D0inverse}
[D^{(0)}_{ab}(\vq,\mathrm{i}\omega_n)]^{-1} = N 
\left(
\begin{array}{llllll}
\frac{\delta^2}{2} \left[ V(\vq)-J(\vq)\right] 
& \frac{\delta}{2} & 0 & 0 & 0 & 0\\
\frac{\delta}{2} & 0 & 0 & 0 & 0 & 0\\
0 & 0 & \frac{4\Delta^2}{J} & 0 & 0 & 0\\
0 & 0 & 0 & \frac{4\Delta^2}{J} & 0 & 0\\
0 & 0 & 0 & 0 & \frac{4\Delta^2}{J} & 0\\
0 & 0 & 0 & 0 & 0 & \frac{4\Delta^2}{J}
\end{array}
\right) \; ,
\end{equation}
\end{widetext}
where $J(\vq) = \frac{J}{2} (\cos q_x +  \cos q_y)$ and the matrix indices $a$ and $b$ run from 1 to 6; 
$\vq$ is a three dimensional wavevector and $\omega_n$ is a bosonic Matsubara frequency. 
This bare propagator $D_{ab}^{(0)}(\vq, {\rm i}\omega_n)$ corresponds to 
the bare charge susceptibilities. 

At leading order, the bare susceptibilities are renormalized to be 
\be
D^{-1}_{ab}(\vq,\mathrm{i}\omega_n)
= [D^{(0)}_{ab}(\vq,\mathrm{i}\omega_n)]^{-1} - \Pi_{ab}(\vq,\mathrm{i}\omega_n)\,,
\label{dyson}
\ee
because of coupling to fermions. The functional forms of the interaction vertices 
can be easily read off from the path integral formalism (see Fig.~1 in Ref.~\onlinecite{foussats04}) and 
is given by the six-component vertex $h_a$ 
\begin{widetext}
\begin{align}
 h_a(\vk,\vq,\nu) =& \left\{
                   \frac{2\varepsilon_{\vk-\vq}+\nu+2\mu}{2}+
                   2\Delta \left[ \cos\left(k_x-\frac{q_x}{2}\right)\cos\left(\frac{q_x}{2}\right) +
                                  \cos\left(k_y-\frac{q_y}{2}\right)\cos\left(\frac{q_y}{2}\right) \right];1;
                 \right. \nonumber \\
               & \left. -2\Delta \cos\left(k_x-\frac{q_x}{2}\right); -2\Delta \cos\left(k_y-\frac{q_y}{2}\right);
                         2\Delta \sin\left(k_x-\frac{q_x}{2}\right);  2\Delta \sin\left(k_y-\frac{q_y}{2}\right)
                 \right\} \, .
\label{vertex-h}
\end{align}
\end{widetext}
The electronic dispersion $\epsilon_{\vk}$ is given by 
\be
\varepsilon_{\vk} = \varepsilon_{\vk}^{\parallel}  + \varepsilon_{\vk}^{\perp} \,,
\label{Ek}
\ee
where the in-plane dispersion $\varepsilon_{\vk}^{\parallel}$ and the out-of-plane dispersion 
$\varepsilon_{\vk}^{\perp}$ are given, respectively, by
\be
\varepsilon_{\vk}^{\parallel} = -2 \left( t \frac{\delta}{2}+\Delta \right) (\cos k_{x}+\cos k_{y})-
4t' \frac{\delta}{2} \cos k_{x} \cos k_{y} - \mu \,,\\
\label{Epara}
\ee
\be
\varepsilon_{\vk}^{\perp} = 2 t_{z} \frac{\delta}{2} (\cos k_x-\cos k_y)^2 \cos k_{z}  \,,
\label{Eperp}
\ee
and $\mu$ is the chemical potential. 
Note that the bare hopping integrals $t$, $t'$, and $t_z$ are renormalized by a factor 
$\delta/2$. This renormalization is a leading order correction due to the 
local constraint because the $t$-$J$ model is defined in the Fock space without double occupancy 
at any site.  
Note that $k_z$ and $q_z$ dependences enter only through $\epsilon_{\vk-\vq}$  in the first column 
in \eq{vertex-h}, whereas the other columns contain only the in-plane momentum $\vq_{\parallel}$.  
Using the vertices $h_a(\vk,\vq,\nu)$, the $6 \times 6$ bosonic self-energies are computed at leading order as 
\begin{eqnarray}
&& \Pi_{ab}(\vq,\mathrm{i}\omega_n)
            = -\frac{N}{N_s N_z}\sum_{\vk} h_a(\vk,\vq,\varepsilon_\vk-\varepsilon_{\vk-\vq}) 
            \frac{n_F(\varepsilon_{\vk-\vq})-n_F(\varepsilon_\vk)}
                                  {\mathrm{i}\omega_n-\varepsilon_\vk+\varepsilon_{\vk-\vq}} 
            h_b(\vk,\vq,\varepsilon_\vk-\varepsilon_{\vk-\vq}) \nonumber \\
&& \hspace{25mm} - \delta_{a\,1} \delta_{b\,1} \frac{N}{N_s N_z}
                                       \sum_\vk \frac{\varepsilon_\vk-\varepsilon_{\vk-\vq}}{2}n_F(\varepsilon_\vk) \; , 
                                       \label{Pi}
\end{eqnarray}
where $N_s$ and $N_z$ are the total 
number of lattice sites on the square lattice and the number of layers along the $z$ direction, respectively. 
Therefore by studying $D_{ab}(\vq,\mathrm{i}\omega_n)$ [\eq{dyson}], we can elucidate all possible 
charge dynamics in the layered $t$-$J$ model at leading order in the large-$N$ expansion.

If $J=0$, $\delta X^a$ reduces to a $2$-component bosonic field $\delta X^a=(\delta R,\delta \lambda)$, 
that is, the bosonic propagator $D_{ab}$ becomes a $2\times2$ matrix and only usual on-site 
charge fluctuations are active. 
When $J$ is finite,  bond-charge fluctuations become active and 
$a$ and $b$ take values from $1$ to $6$. Thus, in principle,  the mixing 
between on-site and bond charge fluctuations is expected for finite $J$. 
We then define three sectors of the $6 \times 6$ bosonic propagator $D_{ab}$: 
a) A $2\times2$ sector for $a,b=1,2$ which 
contains usual on-site charge fluctuations, 
b) a $4\times4$ sector or $J$-sector for $a,b=3$-$6$, 
c) a $2 \times 4$ sector where $a=1,2$ and $b$ runs from $3$ to $6$, and 
a $4 \times 2$ sector with $a=3$-$6$ and $b=1,2$. 

Finally, we mention the connection between the present model and charge excitations 
in the actual CuO$_2$ plane. 
It is known that the $t$-$J$ model is an effective model and each lattice site corresponds to 
a Cu atom in the CuO$_2$ plane. 
However, the $t$-$J$ model is derived from the three-band Hubbard model in the strong 
coupling limit \cite{fczhang88} and thus the effect of O atoms is implicitly considered.
Thus, whereas the usual charge susceptibility describes charge fluctuations 
on each Cu site, the susceptibilities defined by the $J$-sector account for 
charge fluctuations between Cu sites, i.e., on the O sites. 

\section{results}
In the following, we focus on  parameters appropriate for e-cuprates \cite{greco16}, 
$J/t=0.3$, $t'/t=0.30$, and $t_z/t=0.1$.
For cuprates, a bare $t$ is usually assumed to be around $t=500$ meV (Ref.~\onlinecite{hybertsen90}), 
but we assume $t=1$ eV because $t$ is scaled as $t \rightarrow t/N$ in the large-$N$ expansion and $N=2$ 
in the actual material. All quantities with the dimension of energy are measured in units of $t$. 
We take the number of layers as $N_z=30$, which is large enough.
Concerning the long-range Coulomb interaction [\eq{LRC}], we choose 
$d/a=1.5$ (Ref.~\footnote{
Although CuO$_2$ planes shift by $(\frac{1}{2}, \frac{1}{2}, \frac{1}{2})$ in 
Nd$_{2-x}$Ce$_{x}$CuO$_{4}$, we model  
the actual system by neglecting such a shift for simplicity.
Our interlayer distance $d$ is thus given by a half of the $c$-axis lattice constant.
}) ($a=4$~{\AA}), $\epsilon_\parallel=4 \epsilon_0$, and
$\epsilon_\perp=2 \epsilon_0$ where $\epsilon_0$ is the vacuum dielectric constant \cite{timusk89}. 
We set the temperature ($T$) to zero and compute the imaginary part of various charge susceptibilities 
after  analytical continuation ${\rm i}\omega_n \rightarrow \omega + {\rm i} \Gamma$ in \eq{dyson}. 
Here $\Gamma (>0)$ is infinitesimally small and 
we choose $\Gamma = 10^{-3}$ for a numerical convenience.  
Computation is performed mainly for doping  $\delta=0.15$ to make a direct comparison with 
recent experiments in the e-cuprates \cite{ishii14,wslee14,da-silva-neto15,da-silva-neto16}. 
The temperature is $T=0$. 
Since $q_z$ is in general finite in RIXS, we will present results 
mainly for $q_z=\pi$ as representative ones. 

We first study each element of $D_{ab}(\vq,\omega)$ [\eq{dyson}], which has not been 
clarified yet in the literature to the best of our knowledge. 
We then study usual on-site 
charge and bond-charge excitations, which are described by various combinations of 
the elements of $D_{ab}(\vq,\omega)$. 
The superposition of these two kind of charge excitations reveals a dual structure of the charge spectrum. 
We also present results of $\omega$-integrated spectral weight because RXS can measure it directly. 
Finally we make a comparison with results in the standard 2D $t$-$J$-$V$ model.

\subsection{Elements of $\boldsymbol{D_{ab}(\vq,\omega)}$}
\begin{figure}
\centering
\includegraphics[width=16cm,angle=0]{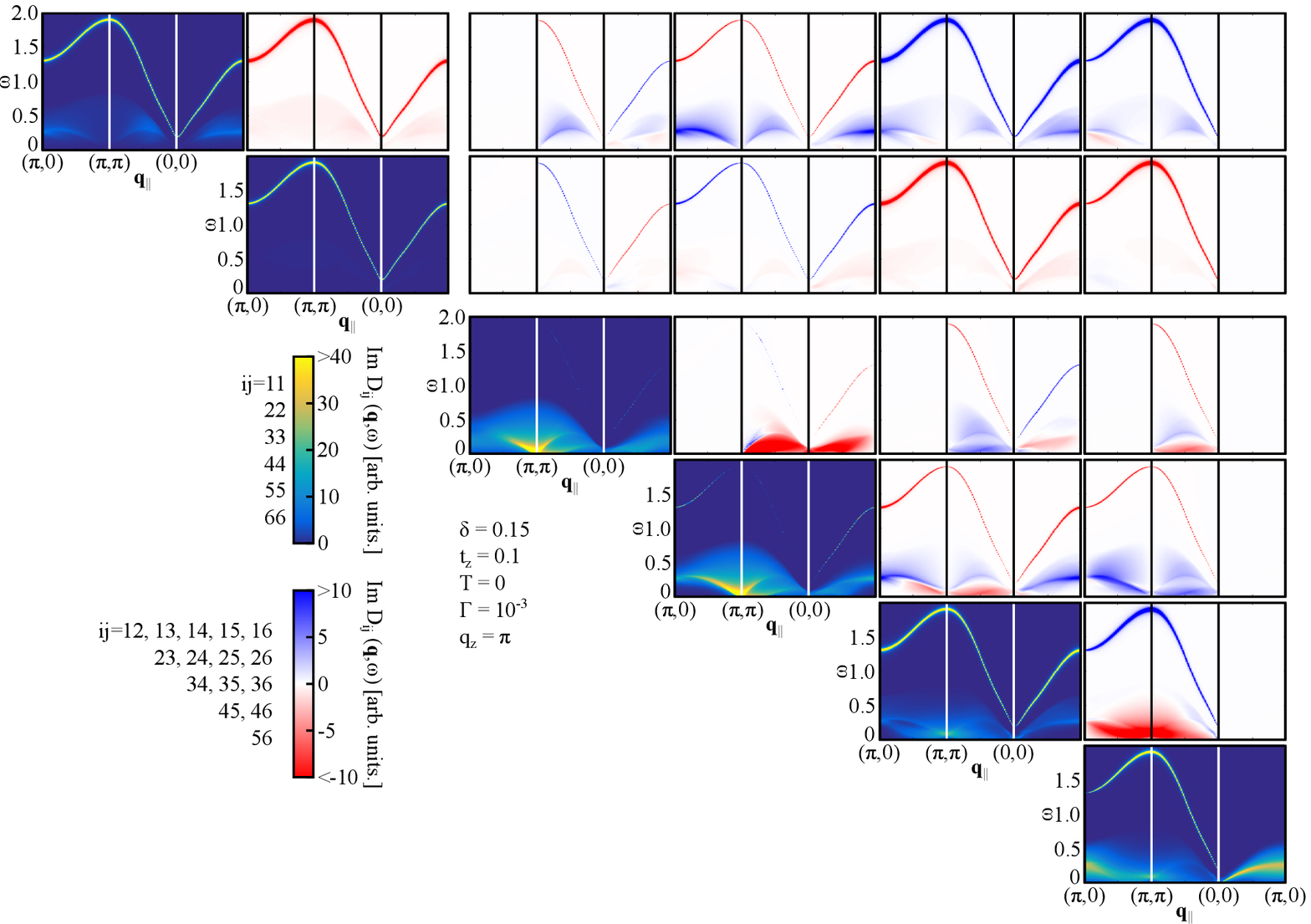}
\caption{(Color online) $\qp$-$\omega$ maps of each element of Im$D_{ab}(\vq,\omega)$ along 
the symmetry axes $(\pi,0)$-$(\pi,\pi)$-$(0,0)$-$(\pi,0)$; 
the out-of-plane momentum is $q_z=\pi$. The elements are placed in a $6 \times 6$ matrix form   
and only the  upper triangle of the matrix is shown because $D_{ab}(\vq,\omega)$ is symmetric. 
To clarify each sector of $2 \times 2$, $4 \times 4$, and $2 \times 4$, a small space is inserted 
between each sector. 
}
\label{qw-map-allD-pi}
\end{figure}

In the present theoretical framework, all possible charge susceptibilities are described by 
various combinations of the elements in the $6\times6$ matrix $D_{ab}(\vq,\omega)$. 
Therefore, we first study excitation spectra of each element of the matrix.
In Fig.~\ref{qw-map-allD-pi}, we show  
$\qp$-$\omega$ maps of Im$D_{ab}({\bf q},\omega)$ along the symmetry axes for $q_z=\pi$. 
Since $D_{ab}(\vq,\omega)$ is a symmetric matrix, we show only the upper triangle of the matrix. 

All diagonal elements of Im$D_{ab}(\vq,\omega)$ have positive spectral weight.  
We see two typical features: i) a continuum spectrum at relatively 
low energy with a scale of $J (=0.3)$ 
and ii) a sharp mode extending from $\omega \approx 0.2$ at 
$\qp=(0,0)$ to $\omega \approx 2.0$ at $\qp=(\pi,\pi)$. 
The former spectrum comes from particle-hole excitations and their spectral intensity in 
$D_{11}$ and $D_{22}$ is much weaker than in $D_{33}$, $D_{44}$, 
and $D_{66}$. 
In particular, the spectral weight of the continuum in $D_{22}$ is almost invisible in the scale of \fig{qw-map-allD-pi}. 
The sharp mode is realized above the particle-hole continuum and originates from 
the zeros of the determinant of $D^{-1}_{ab}(\vq,\omega)$, i.e., it is a collective mode. 
Thus, this mode in principle can appear in all elements 
of $D_{ab}(\vq,\omega)$. 
However, the mode vanishes completely along the $(\pi,0)$-$(\pi,\pi)$ direction in 
$D_{33}$ and along the $(0,0)$-$(\pi,0)$ direction in  $D_{66}$.  
This is because the determinant of $D^{-1}_{ab}(\vq,\omega)$ 
does not affect $D_{33}$ and $D_{66}$ along those directions.
This special feature is connected with 
the vanishing of the spectral weight along the $(\pi,0)$-$(\pi,\pi)$ direction in $D_{13}$, $D_{23}$, 
$D_{34}$, $D_{35}$, $D_{36}$, and along the $(0,0)$-$(\pi,0)$ direction in $D_{a6}$ with $a=1$-$5$ 
(see \fig{qw-map-allD-pi}), which is traced back to the fact that the corresponding  
$\Pi_{ab}(\vq,\omega)$ in \eq{dyson} vanishes due to the symmetry of the vertex $h_a$ [\eq{vertex-h}]. 

The off-diagonal elements of Im$D_{ab}(\vq,\omega)$ might seem odd because some of them 
exhibit {\it negative} spectral weight. Those, of course, do not have a direct physical meaning. 
As we will show explicitly, the physical susceptibilities are defined by some combinations of 
the elements of $D_{ab}(\vq,\omega)$ and they have positive spectral weight. 
Similar to the diagonal part, the off-diagonal elements have two typical features: a
continuum spectrum limited to a low-energy region 
and a collective mode above the continuum. 
This mode is the same as that in the diagonal elements, namely the zeros of 
the determinant of $D^{-1}_{ab}(\vq,\omega)$. 
Another intriguing aspect of \fig{qw-map-allD-pi} is that the continuum spectrum  of $D_{2a}$ with $a=1$-$6$ 
are typically very weak compared with others. 
$D_{2a}$ contains $\delta \lambda$, which describes fluctuations of the Lagrangian 
multiplier imposing non-double occupancy at any site. 
While the case of $q_z=\pi$ is plotted in \fig{qw-map-allD-pi}, the $q_z$ dependence is very weak 
and essentially the same results are obtained for different values of $q_z$ 
except for $q_z \approx 0$ around $\qp=(0,0)$; see Appendix~A and also Ref.~\onlinecite{greco16} 
where the element $(1,1)$ of $D_{ab}(\vq,\omega)$ is studied in detail.

\begin{figure}[t]
\centering
\includegraphics[width=8cm,angle=0]{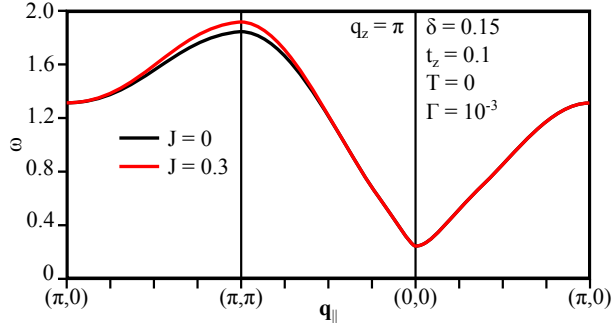}
\caption{(Color online) Zeros
of the determinant of $D_{ab}^{-1}(\vq,\omega)$ along 
the symmetry axes $(\pi,0)$-$(\pi,\pi)$-$(0,0)$-$(\pi,0)$ for $J=0.3$ and $J=0$ at $q_z=\pi$. 
These zeros yield collective charge excitations above the particle-hole continuum. 
}
\label{zero-D-1-pi}
\end{figure}
The collective mode is nothing more than plasmons \cite{greco16}, which are driven by 
the $2\times2$ sector. 
In order to demonstrate this statement, we show in \fig{zero-D-1-pi} the zeros 
of the determinant of $D_{ab}^{-1} (\vq,\omega)$ for $J=0.3$ and $J=0$ along the symmetry axes. 
The zeros appear almost at the same position for both values of $J$. 
Since the $4\times 4$ sector in \eq{dyson} vanishes for $J=0$, we can 
conclude that the collective peaks originate from the $2 \times 2$ sector, namely 
on-site charge  fluctuations.

\subsection{On-site charge susceptibility} 
As discussed previously\cite{foussats04,bejas12}, 
the element $(1,1)$ of $D_{ab}$ is related 
to the usual charge-charge correlation function  
$\chi_c (\vr_i -\vr_j, \tau)=\bra T_\tau n_i(\tau) n_j(0)\ket$, which in the large-$N$ scheme is 
computed in the $\vq$-$\omega$ space as 
\begin{eqnarray}
\chi_{c}(\vq,\omega)= N \left ( \frac{\delta}{2} \right )^{2} D_{11}(\vq,\omega)  \,.
\end{eqnarray}
Thus, 
$\chi_c$ belongs to the $2\times2$ sector. The factor $N$ on the
the right hand side of the expression comes from the definition of the charge correlation function\cite{foussats04} 
in the large-$N$ scheme, and 
cancels the factor $1/N$ coming from $D_{ab}$ [see Eqs.~(\ref{D0inverse}), (\ref{dyson}), and (\ref{Pi})], 
showing that $\chi_c$ is $O(1)$. The factor $(\delta/2)^2$ is due to the fact that the bosonic field $\delta X^a$ 
must be multiplied by $\delta/2$ to becomes the charge fluctuations 
of the original Hubbard operators. 

While phase separation is frequently discussed for 
e-cuprates \cite{bejas14, martins01,macridin06}, 
it does not occur in our model because the long-range Coulomb interaction prevents it. 
Recalling the frustrated phase separation mechanism \cite{emery93}, 
one might expect charge stripes instability triggered by the long-range Coulomb interaction 
near phase separation. This tendency, however, does not occur as studied 
in detail in Ref.~\onlinecite{greco17}. 

\begin{figure}
\centering
\includegraphics[width=8cm,angle=0]{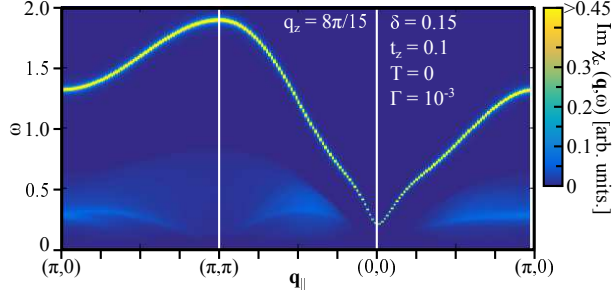}
\caption{(Color online) $\qp$-$\omega$ map of Im$\chi_c(\vq,\omega)$ for $q_z=8\pi/15$. 
$\qp$ is scanned along the symmetry axes $(\pi,0)$-$(\pi,\pi)$-$(0,0)$-$(\pi,0)$. 
}
\label{qw-map-on-site}
\end{figure}

As Im$D_{11}(\vq,\omega)$ is already plotted in \fig{qw-map-allD-pi} for $q_z=\pi$ 
and was studied in detail in Ref.~\onlinecite{greco16}, 
we present results for $q_z=8\pi/15$ in \fig{qw-map-on-site}, which is not seen in the literature 
and thus may be useful for comparison between different values of $q_z$. 
An acoustic-like plasmon mode is clearly seen around $\qp=(0,0)$, which has a gap of $\omega\approx 0.2$
at $\vq_\parallel=(0,0)$. The gap is produced by the finite interlayer hopping $t_z$ (see Fig.~3 in Ref.~\onlinecite{greco16}). 
This plasmon mode with a finite $q_z$ may explain the gapped collective mode observed by 
RIXS (Ref.~\onlinecite{wslee14}). 
A comparison between Fig. 3 and the $(1,1)$ element in \fig{qw-map-allD-pi} demonstrates 
a very weak $q_z$ dependence. However, 
there is a singularity at $q_z=0$ and $\qp=(0,0)$ and we obtain typical 
plasmons with a flat dispersion around  $\qp=(0,0)$ with excitation gap $\omega \approx 0.7$ 
for the present parameters at $q_z=0$ (Ref.~\onlinecite{greco16}) 
(see \fig{qw-map-allD-0} in Appendix~A). This optical plasmon mode can be associated with the plasmons  
observed by optics measurements\cite{singley01} and electron energy-loss 
spectroscopy (EELS) \cite{nuecker89,romberg90}.

\subsection{Bond-charge susceptibilities} 
The study of bond-order instabilities has an old history in the 2D $t$-$J$ model 
\cite{affleck88a,morse91,cappelluti99,foussats04}. 
However, bond-charge susceptibilities as a function of $\qp$ including their dynamics
have not been studied in detail. Neither 
there are attempts of such a study in a realistic model of cuprates such as 
a multilayered $t$-$J$-$V$ model.

For $\delta=0.15$ the homogeneous paramagnetic phase is stable down to $T=0$ 
in the present model. 
However there are various bond-order instabilities in the proximity to this doping that occur
when an eigenvalue of the static $D^{-1}_{ab}(\vq,0)$ vanishes.
The corresponding eigenvector determines the type of instability.
a) $d$-wave bond order \footnote{
Bond orders are referred to as bond-order-phase (BOP) in literatures \cite{bejas12,bejas14}, 
where $d$-wave and $s$-wave bond orders correspond to BOP$_{x\bar{y}}$ and BOP$_{xy}$, respectively.
},
which corresponds to the freezing of the real parts of the bond variable, and the corresponding 
eigenvector is $V^{d{\rm bond}} = \frac{1}{\sqrt{2}} (0,0,1,-1,0,0)$, namely 
the bonds in the $x$ and $y$ directions are in anti-phase [see \eq{bosons}]. 
When the propagation vector is $\qp=(0,0)$, it corresponds to the 
$d$-wave Pomeranchuk instability ($d$PI) \cite{yamase00a,yamase00b,metzner00}. 
b) $s$-wave bond order \cite{Note2}, which 
corresponds also to the freezing of the real parts of the bond variable, but the corresponding eigenvector is 
$V^{s{\rm bond}}= \frac{1}{\sqrt{2}} (0,0,1,1,0,0)$. 
The bonds in the $x$ and $y$ directions are in phase. 
c) As bond orders, 
eigenvectors $V = (0,0,1,0,0,0)$ and $(0,0,0,1,0,0)$ are also possible \cite{bejas12,bejas14} and 
can be described by the superposition of the $d$-wave and $s$-wave bond order.  
d) $d$-wave charge-density wave ($d$CDW) corresponds to the freezing of the imaginary parts of the bond variable, 
and the corresponding eigenvector is 
$V^{d{\rm CDW}} = \frac{1}{\sqrt{2}} (0,0,0,0,1,-1)$. 
When the modulation vector is $\qp = (\pi,\pi)$, it has a $d$-wave character 
and corresponds to the well-known flux phase \cite{affleck88a,morse91,cappelluti99} that describes 
staggered circulating currents.
While charge excitations are usually studied by focusing on a certain type of charge orders, 
we study $d$- and $s$-wave bond orders as well as $d$CDW on an equal footing. 

To define a bond-charge susceptibility, there may be two possibilities: 
i) the projection of $D_{ab}$ onto the corresponding eigenvectors  
$V^{d{\rm bond}}$, $V^{s{\rm bond}}$, and $V^{d{\rm CDW}}$, 
and ii)  the projection of $D_{ab}^{-1}$ onto those eigenvectors. 
Although the former definition might seem natural, that definition in general 
contains the collective mode of the on-site charge fluctuations 
from the $2 \times 2$ sector (see Appendix~B),  which is thus a non-genuine feature of the 
bond-charge fluctuations. On the other hand, 
the contamination from  the on-site charge excitations can be removed by adopting the latter definition. 
Therefore we define each bond-charge susceptibility as 
$\chi_{d{\rm bond}}^{-1} (\vq,\omega)=(1/N)(\delta/2)^{-2}(D_{33}^{-1}+D_{44}^{-1}-2 D_{34}^{-1})/2$, 
$\chi_{s{\rm bond}}^{-1}(\vq, \omega)=(1/N)(\delta/2)^{-2}(D_{33}^{-1}+D_{44}^{-1}+2 D_{34}^{-1})/2$, 
and $\chi_{d{\rm CDW}}^{-1}(\vq, \omega)=(1/N)(\delta/2)^{-2}(D_{55}^{-1}+D_{66}^{-1} -2 D_{56}^{-1})/2$; 
note that $\chi_{d{\rm bond}}$ is referred to as $\chi_{d{\rm PI}}$ in Ref.~\onlinecite{bejas12} 
and $\chi_d$ in Ref.~\onlinecite{yamase15b}. 
The difference between $\chi_{d{\rm bond}}$ and $\chi_{s{\rm bond}}$ lies in the sign 
in front of $D_{34}^{-1}$. As a result, both quantities become identical for $\qp=(\pi, q_y)$ and $(q_x,\pi)$. 
This is easily understood. We have $h_3=-2\Delta \sin k_x$ and 
$h_4=-2\Delta \cos(k_y-\frac{q_y}{2})$ for $\qp=(\pi,q_y)$ in \eq{vertex-h} and thus 
$\Pi_{34}$ in \eq{Pi} vanishes for $\qp=(\pi, q_y)$ and $(q_x,\pi)$, 
leading to $D_{34}^{-1}=0$ there [see Eqs.~(\ref{D0inverse}) and (\ref{dyson})].

\begin{figure}
\centering
\includegraphics[width=8cm]{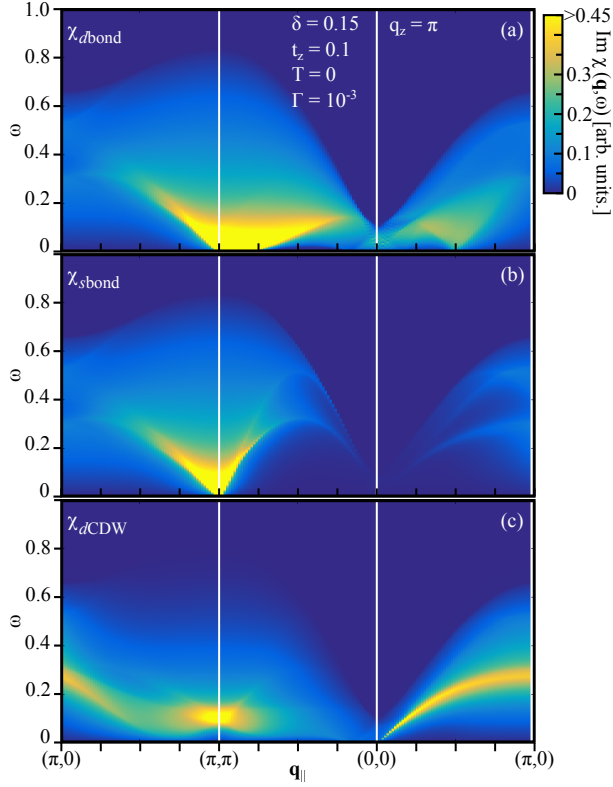}
\caption{(Color online) $\vq_{\parallel}$-$\omega$ maps of the spectral weight for 
(a) $\chi_{d{\rm bond}}$, 
(b) $\chi_{s{\rm bond}}$, and (c) $\chi_{d{\rm CDW}}$ along the symmetry axes $(\pi,0)$-$(\pi,\pi)$-$(0,0)$-$(\pi,0)$.
The out-of-plane momentum is $q_z=\pi$.}
\label{qw-map-D-1-pi}
\end{figure}

Bond-charge excitation spectra of $\chi_{d{\rm bond}}(\vq,\omega)$, 
$\chi_{s{\rm bond}}(\vq,\omega)$, and  $\chi_{d{\rm CDW}}(\vq,\omega)$ are shown in 
\fig{qw-map-D-1-pi} along the symmetry axes for $q_z=\pi$. 
Before going into details of each excitation spectrum, we first discuss the overall features 
in \fig{qw-map-D-1-pi}. 
First of all, all susceptibilities exhibit positive spectral weight, although 
some elements of $D_{ab}(\vq,\omega)$ contain negative spectral weight in 
\fig{qw-map-allD-pi}.  
Since our choice of doping $\delta=0.15$ is close to CO instabilities, large spectral weight is present 
in a relatively low-energy region $\omega \lesssim 0.2$. 
The spectral weight in a high energy region is very broad and diffusive.  
Those features are essentially 
independent of $q_z$, and very similar results are obtained even in the 2D case (see Sec.~III F). 

$\chi_{d{\rm bond}}$ shows large spectral weight at low energy around $\vq_{\parallel}=0.8(\pi,\pi)$.
This spectral weight is associated with the leading soft mode at $\vq_\parallel \approx 0.8(\pi,\pi)$, 
which accumulates spectral weight upon approaching the CO instability at $\delta_c = 0.129$.
Along the direction $(0,0)$-$(\pi,0)$ the spectrum has rather high intensity and its energy goes down 
toward the momentum $\vq_\parallel=(0.5\pi,0)$. 
This subleading mode may correspond to the CO features observed in RXS experiments \cite{da-silva-neto15}, 
as was discussed in Ref.~\onlinecite{yamase15b}; see also Sec.~III E and Sec.~IV.

$\chi_{s{\rm bond}}$ shows a low-energy dispersion around $\vq_\parallel=(\pi,\pi)$, which is 
related to the proximity to the corresponding instability at 
$\delta_c = 0.111$. 
Its spectral weight disperses upwards forming a V-shape and loses intensity with increasing 
$\omega$. In contrast to the case of $\chi_{d{\rm bond}}$, 
there is no CO tendency along $(0,0)$-$(\pi,0)$ direction. 
Instead, along that direction, there is a very weak dispersive feature, which reaches 
$\omega \approx 0.2$ at $\qp=(0.5\pi,0)$. This reflects a subtle structure of 
individual particle-hole excitations.
 
$\chi_{d{\rm CDW}}$ exhibits large spectral weight at $\vq_\parallel =(\pi,\pi)$ around $\omega = 0.1$. 
This energy is reduced to zero with decreasing doping towards 
$\delta_c=0.093$, where the $d$CDW instability occurs. 
Interestingly, there is a clear gapless 
dispersion along $(0,0)$-$(\pi,0)$ direction and it extends up 
to $\omega \sim 0.3$ at $\vq_\parallel=(\pi,0)$. 
This is not a collective mode associated with the $d$CDW, but a peak structure of individual 
excitations that originates from a local minimum of the real part of the denominator 
of $\chi_{d{\rm CDW}}$. The dispersive feature is not clearly seen along $(0,0)$-$(\pi,\pi)$ direction, 
showing an asymmetric character of $\chi_{d{\rm CDW}}$. 
We also see the dispersive feature in the $(\pi,0)$-$(\pi,\pi)$ direction, which 
merges into the large spectral weight at $\vq_\parallel=(\pi,\pi)$ and 
$\omega \approx  0.1$. 

All of these bond-charge susceptibilities originate from the $4 \times 4$ sector, 
namely $J$-sector in \eq{dyson}. 
Hence one might assume that a magnetic probe should be used for 
their detection. However, a charge probe such as RIXS can, in principle, be used for testing our results,
even in the case of the $d$CDW where most of discussions focus on the possibility for detecting the small magnetic 
signal corresponding to the circulating currents.

\subsection{Dual structure of charge-excitation spectrum} 
We have presented separately usual on-site charge excitations (Sec.~III~B) and 
various bond-charge  excitations (Sec.~III~C). 
While it is not clear whether RIXS can detect 
them in a selective way or as a sum with a certain weight for each susceptibility, 
the full charge excitation spectrum in e-cuprates may be described by the superposition of 
all possible charge excitations. 

\begin{figure}
\centering
\includegraphics[width=8cm]{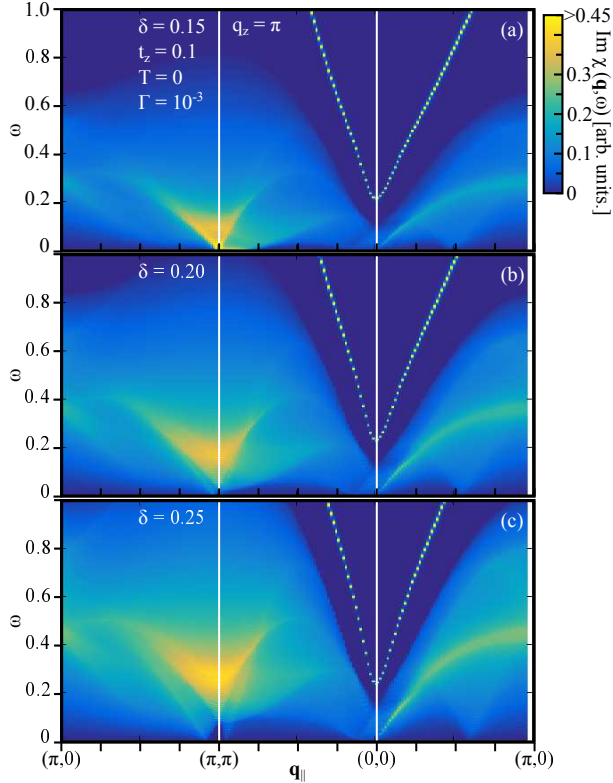}
\caption{(Color online) $\qp$-$\omega$ maps of the superposition of excitation spectra of 
$\chi_c$, $\chi_{d{\rm bond}}$, $\chi_{s{\rm bond}}$, and  $\chi_{d{\rm CDW}}$ 
along the symmetry axes $(\pi,0)$-$(\pi,\pi)$-$(0,0)$-$(\pi,0)$ for $q_z=\pi$ at three different doping rates: 
(a) $\delta=0.15$, (b) $\delta=0.20$, and (c) $\delta=0.25$.}
\label{qw-map-sum-pi}
\end{figure}

Figure~\ref{qw-map-sum-pi}(a) shows the superposition of the spectra of 
$\chi_c$, $\chi_{d{\rm bond}}$, $\chi_{s{\rm bond}}$, and  $\chi_{d{\rm CDW}}$ for $q_z=\pi$. 
It turns out that the {\it total} charge excitation spectrum is characterized by a dual structure: 
a gapped V-shaped dispersion around $\vq=(0,0)$ extending to a high-energy 
region and a continuum spectrum limited to a low-energy region typically 
with a scale of $J (=0.3)$. 
As we have shown in Sec.~III~A and B, the V-shaped dispersion originates from the collective 
on-site charge excitations, namely a plasmon mode with a finite $q_z$. 
This mode is controlled exclusively by the $2 \times 2$ sector in \eq{dyson}, with no 
contribution from the bond-charge excitations. 
On the other hand, the low-energy spectrum is the superposition 
of $\chi_{d{\rm bond}}$, $\chi_{s{\rm bond}}$, and $\chi_{d{\rm CDW}}$, and the particle-hole 
continuum from $\chi_c$. 
The former three contributions have much larger spectral weight than  the last one. 
In \fig{qw-map-sum-pi}(a), we can see a large spectral weight only around $\qp=(\pi,\pi)$ 
in a very low-energy region. Unfortunately RIXS cannot explore a region near $\qp=(\pi,\pi)$. 
The subleading CO tendency occurring at $\qp=(0.5\pi,0)$ [see \fig{qw-map-D-1-pi}(a)] 
becomes unclear when the spectra are superimposed in \fig{qw-map-sum-pi}(a), 
but a cusplike feature of the continuum at $\qp=(0.5\pi,0)$ and $\omega=0$ is barely discernible. 

The dual structure should not be mixed with the usual distinction between collective 
excitations and the continuum spectrum. The intriguing aspect here is that their origins  
are completely different: the former comes from usual on-site charge fluctuations 
and the latter from bond-charge fluctuations. Furthermore the continuum spectra are not 
composed of simply a certain type of charge order, but of various types of bond-charge fluctuations 
such as $d$-wave bond, $s$-wave bond, and $d$CDW. 

Because of the dual structure of charge excitations, their doping dependence becomes distinct. 
Figures~\ref{qw-map-sum-pi}(b) and (c) show results for higher doping. 
The plasmon mode remains clear and nearly unchanged with increasing doping. 
On the other hand, the continuum spectrum typically becomes broader, but 
some characteristic features occur. First,   
the low-energy spectrum around $\qp=(\pi,\pi)$ 
hardens with increasing doping. This feature is easily expected because 
the system goes further away from the critical doping of the bond-order instability.  
However, unexpectedly the spectral intensity remains rather strong even at high doping. This 
is due to the factor $(\delta/2)^{2}$ in front of bond-charge susceptibility, a specific aspect 
of strong correlation effects in the $t$-$J$ model; 
see also the factor $\delta/2$ in Eqs.~(\ref{Epara}) and (\ref{Eperp}). 
Second, along the direction $(0,0)$-$(\pi,0)$, a peak of the continuum becomes clearer with increasing doping. 
This peak originates from individual fluctuations 
associated with $d$CDW as seen in \fig{qw-map-D-1-pi}(c).

\subsection{$\boldsymbol{\omega}$-integrated spectral weight $\boldsymbol{S(\vq)}$} 
While we have presented $\omega$-{\it resolved} spectral weight, 
it is also interesting to study the $\omega$-{\it integrated} spectral weight for each $\vq$, 
namely the equal-time correlation function $S(\vq)$, because RXS can measure it directly. 
For a given susceptibility $\chi$, it is defined as 
\be
S(\vq)=\frac{1}{\pi} \int_{-\infty}^\infty {\rm d}\omega {\rm Im} \chi(\vq,\omega)
[n_B(\omega)+1]
\label{eqSq}
\ee
where $n_B$ is the Bose factor.

\begin{figure}
\centering
\includegraphics[width=8cm]{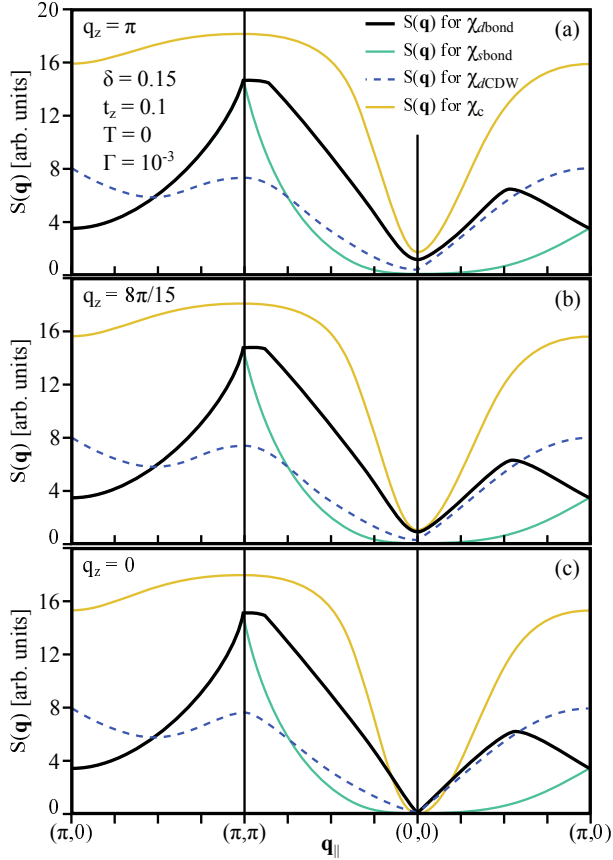}
\caption{(Color online) $S(\vq)$ at $\delta=0.15$ and $T=0$ for
$\chi_{d{\rm bond}}$, $\chi_{s{\rm bond}}$, $\chi_{d{\rm CDW}}$, and $\chi_c$ 
along the symmetry axes $(\pi,0)$-$(\pi,\pi)$-$(0,0)$-$(\pi,0)$ 
for (a) $q_z=\pi$, (b) $q_z=8\pi/15$, and (c) $q_z=0$. 
}
\label{Sq}
\end{figure}

In \fig{Sq}(a) we show $S(\vq)$ for
$\chi_{d{\rm bond}}$, $\chi_{s{\rm bond}}$,
$\chi_{d{\rm CDW}}$, and $\chi_c$ along the symmetry axes at $q_z=\pi$ and $\delta=0.15$ at $T=0$.
The obtained results are easily understood by referring to \fig{qw-map-D-1-pi}. The spectral 
weight of Im$\chi_{d{\rm bond}}$ is very large around $\qp=(\pi,\pi)$. Thus $S(\vq)$ also exhibits 
a peak structure around $\qp=(\pi,\pi)$ with large weight to the side of $(\pi,\pi)$-$(0,0)$. 
Importantly, there is a sub-leading peak at $\vq_\parallel=(0.5\pi,0)$ along the axial direction 
$(0,0)$-$(\pi,0)$. This sub-leading peak is present only for $d$-wave bond fluctuations and originates 
from the low-energy spectral weight at $\vq_\parallel=(0.5\pi,0)$ in \fig{qw-map-D-1-pi}(a). 

While a curve for $s$-wave bond fluctuations is not seen along the direction $(\pi,0)$-$(\pi,\pi)$ 
in \fig{Sq}, $S(\vq)$ for $s$-wave bond fluctuations is identical to that for $d$-wave bond fluctuations 
along that direction because of symmetry (see Sec.~III C). 
As expected from \fig{qw-map-D-1-pi}(b), $S(\vq)$ for $s$-wave bond exhibits a dominant peak 
at $\qp=(\pi,\pi)$. 
$S(\vq)$ for $d$CDW also exhibits a peak at $\qp=(\pi,\pi)$ much broader than that for $s$-wave 
bond. Interestingly, there is another broad peak at $\qp=(\pi,0)$, which comes from the 
spectral weight associated with the dispersive feature along $(0,0)$-$(\pi,0)$-$(\pi,\pi/2)$ 
in \fig{qw-map-D-1-pi}(c). 
For on-site charge fluctuations, $S(\vq)$ exhibits a broad dip structure around $\qp=(0,0)$. 
This is because the corresponding Im$\chi_c(\vq,\omega)$ [see the $(1,1)$ element in 
\fig{qw-map-allD-pi} and \fig{qw-map-on-site}] has lower intensity 
around $\qp=(0,0)$ due to a small phase space of particle-hole excitations. 

For completeness, we also present results for different values of $q_z$ in Figs.~\ref{Sq}(b) and (c). 
These results are essentially the same as \fig{Sq}(a) at $q_z=\pi$, indicating very weak $q_z$ dependence 
of $S(\vq)$. Quantitative changes occur around $\qp=(0,0)$, where the spectral weight 
vanishes for all $\chi_{d{\rm bond}}$, $\chi_{s{\rm bond}}$, $\chi_{d{\rm CDW}}$, and $\chi_{c}$ 
at $q_z=0$ [\fig{Sq}(c)], because Im$\chi$ vanishes at $\vq=(0,0,0)$.

\subsection{Two-dimensional $\boldsymbol{t}$-$\boldsymbol{J}$-$\boldsymbol{V}$ model}
In the present study we have employed a layered $t$-$J$-$V$ model with the long-range 
Coulomb interaction. In the literature, however, most of theoretical studies employ 
a 2D model. Sometimes the effect of the Coulomb interaction is 
also studied, but it is usually limited to the nearest-neighbor interaction such as 
$V(\vq)=V(\cos q_x+\cos q_y )$. Hence it is useful to study  
the standard 2D $t$-$J$-$V$ model with the nearest-neighbor Coulomb interaction and 
clarify the differences with previous results. 
We set $t_z=0$ in \eq{Eperp} and replace \eq{LRC} by $V(\vq)=V(\cos q_x+\cos q_y)$ with $V=1$, 
otherwise we employ the same parameters. 

\begin{figure}[t]
\centering
\includegraphics[width=8cm]{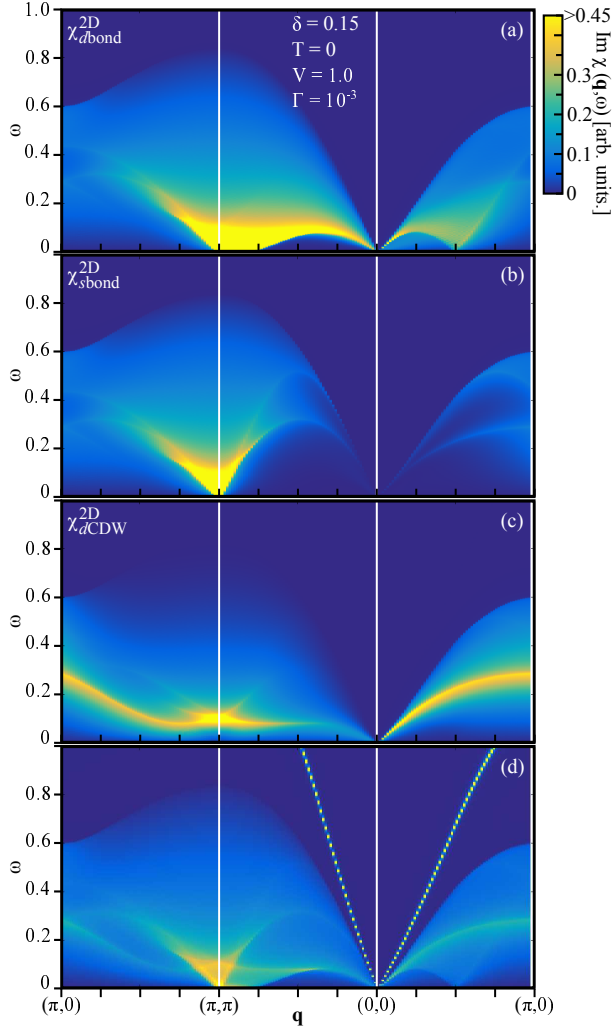}
\caption{(Color online) Charge excitation spectra for (a) $\chi^{\rm 2D}_{d{\rm bond}}$, 
(b) $\chi^{\rm 2D}_{s{\rm bond}}$, and (c) $\chi^{\rm 2D}_{d{\rm CDW}}$ in the 2D $t$-$J$-$V$ model 
along the symmetry axes $(\pi,0)$-$(\pi,\pi)$-$(0,0)$-$(\pi,0)$. 
(d) Superposition of the spectra of $\chi^{\rm 2D}_{c}$, 
$\chi^{\rm 2D}_{d{\rm bond}}$,  $\chi^{\rm 2D}_{s{\rm bond}}$, and $\chi^{\rm 2D}_{d{\rm CDW}}$. 
}
\label{qw-map-2D}
\end{figure}

In Figs.~\ref{qw-map-2D}(a)-(c), we show results for $\chi^{\rm 2D}_{d{\rm bond}}$, 
$\chi^{\rm 2D}_{s{\rm bond}}$, and $\chi^{\rm 2D}_{d{\rm CDW}}$, respectively. 
A comparison with \fig{qw-map-D-1-pi} shows that the distribution of the spectral weight 
is very similar to each other. 
This means that bond-charge fluctuations are intrinsically of 2D character and are controlled 
by the $J$-term of the CuO$_2$ plane. Although the functional form of $V(\vq)$ 
used in Figs.~\ref{qw-map-D-1-pi} and \ref{qw-map-2D}
is very different, such an impact is minor on bond-charge fluctuations. 
This is easily understood.  
In the bare bosonic propagator [\eq{D0inverse}], 
$V(\vq)$ enters the $2 \times 2$ sector whereas the bond field $\Delta$ the $4 \times 4$ sector.  
In principle, both sectors interfere with each other through \eq{dyson}. 
However, our actual calculations in Sec.~III A-D have showed that both sectors are essentially 
decoupled. 
In addition, we find that the critical doping and critical 
temperatures of bond orders in the 2D model \cite{bejas14} are almost the same as 
those obtained in the layered model. This is another evidence that bond-charge fluctuations are 
controlled by the $J$-term in the 2D plane. 

Usual on-site charge excitations ($\chi_{c}^{\rm 2D}$) were 
already computed in Fig.~1(c) in Ref.~\onlinecite{greco17}. 
Instead of reproducing such a result, we show the superposition of 
Im$\chi^{\rm 2D}_{c}$, Im$\chi^{\rm 2D}_{d{\rm bond}}$, Im$\chi^{\rm 2D}_{s{\rm bond}}$, and 
Im$\chi^{\rm 2D}_{d{\rm CDW}}$ in \fig{qw-map-2D}(d). Similar to the case of \fig{qw-map-sum-pi}, 
the continuum spectra come mainly from bond-charge fluctuations. The major contribution from the 
on-site charge excitations is a gapless linear dispersion around $\vq=(0,0)$ above the continuum. 
This is not a plasmon mode, but a zero-sound mode due to  the short-range Coulomb interaction $V(\vq)$. 
Hence the spectrum of Im$\chi_c^{\rm 2D}$ 
is very different from Sec.~III B, where $q_z$ is finite and 
we have obtained plasmon excitations with 
an energy gap which is scaled by interlayer hopping $t_z$ (see Fig.~3 in Ref.~\onlinecite{greco16}).  
Hence the functional form of $V(\vq)$ and dimensionality are crucial to the collective mode 
above the particle-hole continuum. 

Since the form of $V(\vq)$ is important to on-site charge excitations, it is insightful to mention also 
results in the presence of the long-range Coulomb interaction. 
Typically we may have two cases. 
i) The 2D $t$-$J$ model with  the three dimensional long-range Coulomb interaction 
[\eq{LRC}]. While for $q_z=0$ the usual optical plasmon mode is realized around $\qp=(0,0)$ as seen in the $(1,1)$ element 
in \fig{qw-map-allD-0},  the plasmon 
energy drops to zero once $q_z$ becomes finite and a gapless acoustic plasmon mode is realized. 
This corresponds to the present model with $t_z=0$ [see Fig. 3 in Ref.~\onlinecite{greco16}] and 
to the case of Ref.~\onlinecite{prelovsek99}. 
ii) The 2D $t$-$J$ model with the 2D long-range Coulomb interaction, i.e., $V(\vq) \sim 1/q$. In this case, 
we obtain a gapless plasmon dispersion $\omega(\vq) \sim \sqrt{q}$ as is well known \cite{giuliani}. 

\section{Discussions}
We have shown that the charge excitation spectrum in the $t$-$J$ model exhibits 
a dual structure. One is usual on-site charge fluctuations, which are realized 
as plasmons in a relatively high-energy region ($\omega \gtrsim 0.3$)
above the particle-hole continuum.
The other is bond-charge fluctuations driven by the exchange interaction $J$ and can 
lead to CO phenomena. Its spectral weight appears in a low-energy region, 
typically less than $J (=0.3)$. 
Since the lattice site in the present model corresponds to a Cu atom in the CuO$_{2}$ plane, 
we reasonably interpret that the on-site charge fluctuations correspond to fluctuations 
on the Cu sites and bond-charge fluctuations, i.e., fluctuations between the nearest-neighbor Cu sites 
correspond to charge fluctuations at the O sites. 
This interpretation is actually supported by explicit calculations \cite{atkinson15} in the 
three-band Hubbard model where both Cu and O sites are included. 
Hence  
our theory suggests that the high-energy charge fluctuations consist mainly of the on-site charge 
fluctuations at the Cu sites, and low-energy charge excitations originate mainly from 
the O sites.

The high-energy charge excitations are captured correctly only when both the interlayer coupling 
and the long-range Coulomb interaction are taken into account. 
The 2D $t$-$J$ model studied frequently cannot capture 
the correct high-energy features even if the long-range Coulomb repulsion is included. 
On the other hand, the low-energy features can be well described by the 2D $t$-$J$ model 
and the long-range Coulomb interaction is unimportant. 
The distinction between low- and high-energy regimes, which we call a dual structure, is 
crucial  to understand the complicated charge excitations in cuprates. 
Theoretically this dual structure means, in the language of the $6 \times 6$ bosonic propagator $D_{ab}$ [\eq{dyson}], 
that the $2 \times 2$ sector and  the $4 \times 4$ sector or $J$-sector 
are essentially decoupled; the former controls on-site charge fluctuations and the latter bond-charge fluctuations.

The experimental indications of the underlying dual structure of charge excitation spectrum  
are actually obtained by different methods. 
RIXS revealed the quasi-linear dispersive mode near $\qp=(0,0)$ (Refs.~\onlinecite{wslee14} and \onlinecite{ishii14}).  
Since the value of $q_z$ is finite in RIXS, this mode is interpreted as an acoustic-like 
plasmon mode from the usual on-site charge excitations \cite{greco16}. 
On the other hand, charge excitations were also reported by RXS \cite{da-silva-neto15,da-silva-neto16}. 
RXS measures $\omega$-integrated spectral 
weight for each $\vq$ [\eq{eqSq}] and reported a short-range charge order with 
$\qp \approx (0.48 \pi,0)$ in e-cuprates \cite{da-silva-neto15,da-silva-neto16}. 
This signal was discussed as originating from
$\chi_{d{\rm bond}}$ (Ref.~\onlinecite{yamase15b}), namely from 
the low-energy charge fluctuations driven by the $J$-term. 
While the 2D $t$-$J$ model was employed in Ref.~\onlinecite{yamase15b}, the present work (\fig{Sq}) 
indeed finds a peak near $\qp=(0.5\pi,0)$ only in the $d$-wave bond-charge susceptibility,  
confirming that the conclusions obtained in Ref.~\onlinecite{yamase15b} 
are valid also in the layered $t$-$J$-$V$ model.  
Therefore the short-range charge-order observed in e-cuprates \cite{da-silva-neto15,da-silva-neto16} 
can be a short-range $d$-wave bond order. 

An additional implication from the present work is that there 
can be a stronger tendency of bond-charge orders near $\qp=(\pi,\pi)$ as seen in \fig{qw-map-D-1-pi}. 
Interestingly not only $\chi_{d{\rm bond}}$ but also $\chi_{s{\rm bond}}$ and 
$\chi_{d{\rm CDW}}$ contribute to the low-energy charge fluctuations around 
$\qp=(\pi,\pi)$. Unfortunately this possibility cannot be tested 
by resonant x-ray measurement because such a momentum region is out of its range. 
Further experimental techniques are necessary for detecting them. 
Far away from $(\pi,\pi)$, on the other hand, $\chi_{d{\rm bond}}$ and $\chi_{d{\rm CDW}}$ exhibit 
a dispersive feature along $(0,0)$-$(\pi,0)$ direction 
[see Figs.~\ref{qw-map-D-1-pi}(a), \ref{qw-map-D-1-pi}(c), \ref{qw-map-2D}(a), and \ref{qw-map-2D}(c)], 
which may be tested by RIXS.

As clarified in Ref.~\onlinecite{bejas14}, the tendency to CO instabilities in e-cuprates 
is very different from that in h-cuprates. 
Hence we expect strong particle-hole asymmetry of bond-charge fluctuations driven by 
the exchange interaction $J$. The present results of bond-charge excitation spectra 
may not be applied straightforwardly to h-cuprates, but provide a good basis to understand 
complicated charge excitation spectra. On the other hand, the usual on-site charge excitations, 
namely plasmon excitations, are general features and we expect a similar 
mode also in h-cuprates. In fact, such a mode with a finite $q_z$ 
seems to be observed in RIXS in Ref.~\onlinecite{ishii17}. 
However, the authors of Ref.~\onlinecite{ishii17} 
interpreted them as individual on-site charge excitations, not as plasmons. 
In the present work, the individual on-site charge excitations, namely the continuum excitations, 
are weak as seen in the low-energy region in Fig. 1. Moreover, its spectral weight is even weaker than 
that of bond-charge excitations. Hence plasmons seem a more natural interpretation 
of the data of Ref.~\onlinecite{ishii17}. 

While $q_z$ is finite in RIXS, the plasmon mode with $q_z=0$ can be detected in EELS  \cite{nuecker89,romberg90}. 
EELS observes the loss function, namely Im$(\frac{1}{\epsilon(\vq,\omega)}) \propto V(\vq){\rm Im}\chi_c(\vq,\omega)$. 
The spectral weight of Im$\chi_c(\vq,\omega)$  is proportional to $q^2$ at the plasmon energy and thus 
vanishes at $\vq=(0,0,0)$ [see \fig{Sq}(c) and the (1,1) element in \fig{qw-map-allD-0}]. However, 
the loss function becomes finite at $\vq=(0,0,0)$ because of the singularity of the long-range Coulomb interaction 
in the long-wavelength limit, namely $V(\vq) \propto q^{-2}$. 

\section{Conclusions}
The layered $t$-$J$-$V$ model treated in a large-$N$ expansion shows a dual
structure of charge excitation spectra. 
i) In a low-energy region, the spectral weight originates mainly from various types of 
bond-charge fluctuations with internal symmetry, for example, $\cos k_x -\cos k_y$. 
These bond-charge fluctuations are triggered by the exchange term $J$, and 
can lead to the charge order phenomena. 
The low-energy spectral weight is essentially independent of the out-of-plane momentum $q_z$. 
ii) In the high-energy region, the spectral weight is dominated by usual on-site charge
fluctuations and is characterized by a plasmon mode with a minimal excitation gap 
at $\qp=(0,0)$. The dispersion around $\qp=(0,0)$ is quasi-linear for a
finite $q_z$ whereas it is flat for $q_z=0$ similar to usual optical plasmons. 
The dual structure of charge excitation spectra that we have elucidated in the present paper 
will serve to disentangle major charge excitations from complicated spectra of RIXS data.

\acknowledgments
The authors thank K. Ishii and T. Tohyama for very fruitful discussions about RIXS. 
H.Y. acknowledges support by JSPS KAKENHI Grant No.~15K05189. 
A.G acknowledges the Japan Society for the Promotion of Science
for a Short-term Invitational Fellowship program (S17027) under which this work was completed.

\appendix

\section{Elements of Im$\boldsymbol{D_{ab}(\vq,\omega)}$ at different $\boldsymbol{q_z}$} 
\begin{figure}
\centering
\includegraphics[width=16cm,angle=0]{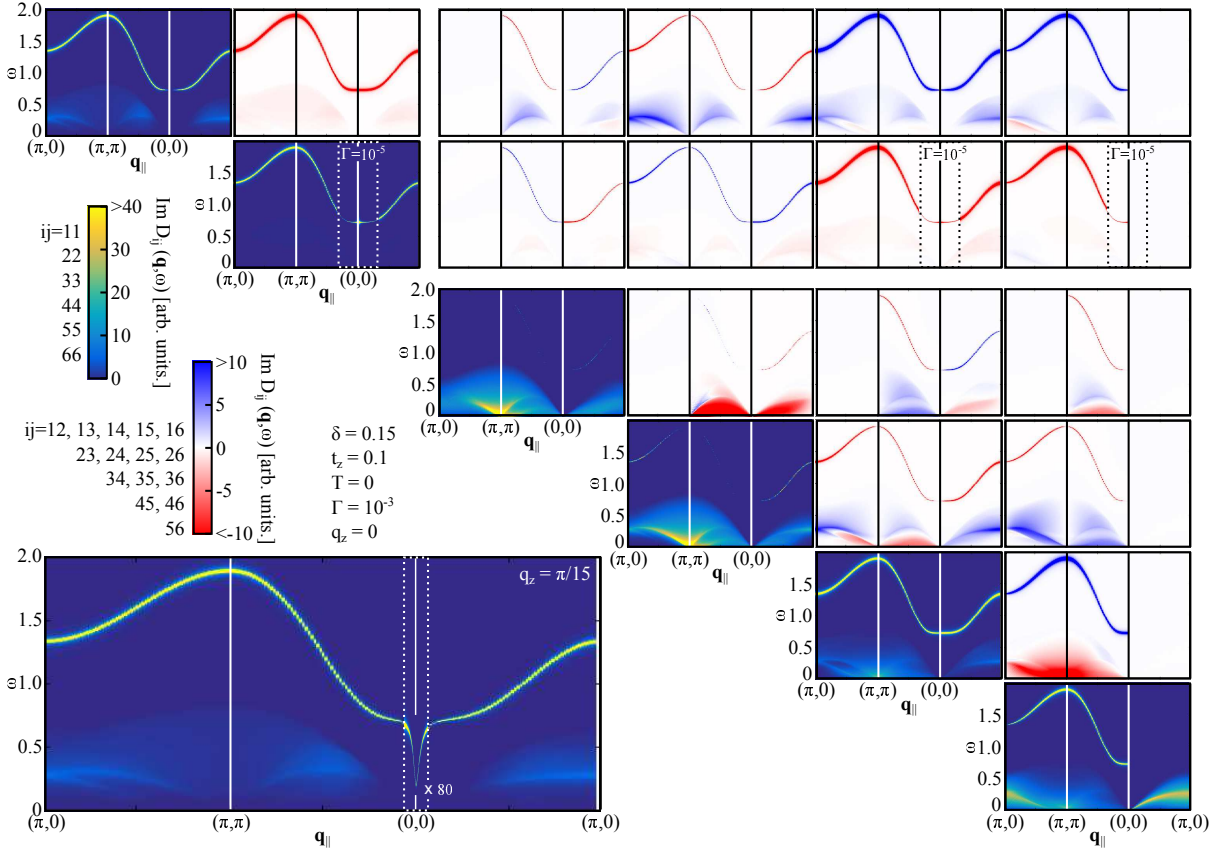}
\caption{(Color online) $\qp$-$\omega$ maps of each element of Im$D_{ab}(\vq,\omega)$ along the symmetry axes $(\pi,0)$-$(\pi,\pi)$-$(0,0)$-$(\pi,0)$. 
The out-of-plane momentum is $q_z=0$. The elements are placed in a $6 \times 6$ matrix form  
and only the  upper triangular matrix is shown because $D_{ab}(\vq,\omega)$ is symmetric. 
To clarify each sector of $2 \times 2$, $4 \times 4$, and $2 \times 4$, a small space is inserted 
between each sector. In the elements $(2,2)$, $(2,5)$, and $(2,6)$, we choose a smaller 
$\Gamma=10^{-5}$ to increase the accuracy when computing near $\vq_\parallel=(0,0)$. 
The inset (bottom left) is the element $(1,1)$ at $q_z=\pi/15$. To make a better contrast, 
the spectral intensity around $\vq_\parallel=(0,0)$ is multiplied by 80 and the vertical  axis line at $\vq_\parallel=(0,0)$ is broken 
between $\omega=0.1$ and $0.75$. 
}
\label{qw-map-allD-0}
\end{figure}

As we have mentioned in Sec.~III~A, each element of Im$D_{ab}(\vq,\omega)$ exhibits 
essentially the same result for different values of $q_z$ except for the collective mode around $\qp=(0,0)$.  
To demonstrate such a statement, we present results for $q_z=0$ in \fig{qw-map-allD-0}. 
A comparison with the result for $q_z=\pi$ (\fig{qw-map-allD-pi}) shows that the continuum spectra 
are in fact essentially the same and the crucial difference is recognized in the plasmon mode around $\qp=(0,0)$. 
The plasmons feature a flat dispersion around $\qp=(0,0)$ for $q_z=0$ and 
a quasi-linear dispersion for $q_z=\pi$. The flat dispersion, however, appears only at $q_z=0$ 
and its gap suddenly drops from $\omega \approx 0.7$ to $\omega \approx 0.2$ once $q_z$ becomes finite, 
yielding a sharp V-shaped dispersion as shown in the inset of \fig{qw-map-allD-0}. 
With increasing $q_z$, the energy gap stays at $\omega \approx 0.2$ and the slope of the V-shaped 
dispersion decreases. For $q_z \gtrsim 0.4\pi$, the V-shape dispersion becomes almost the same 
as that at $q_z=\pi$, which was clearly shown in Fig.~2 in Ref.~\onlinecite{greco16}. 

\section{Alternative definition of bond-charge susceptibility} 
\begin{figure}
\centering
\includegraphics[width=8cm]{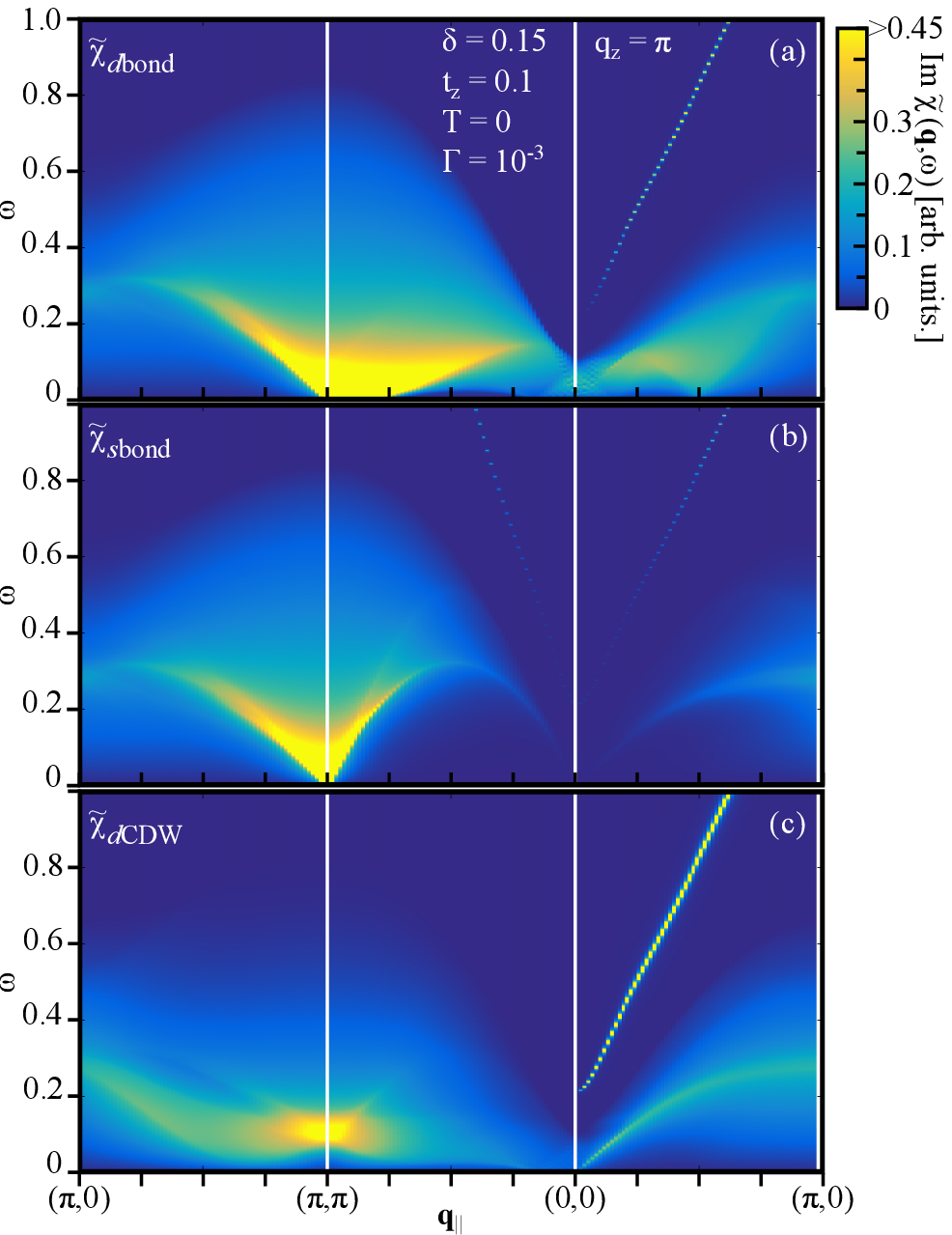}
\caption{(Color online) $\vq_{\parallel}$-$\omega$ maps of Im$\tilde{\chi}_{d{\rm bond}} (\vq,\omega)$, 
Im$\tilde{\chi}_{s{\rm bond}}(\vq,\omega)$, and Im$\tilde{\chi}_{d{\rm CDW}}(\vq,\omega)$ 
along the symmetry axes $(\pi,0)$-$(\pi,\pi)$-$(0,0)$-$(\pi,0)$; 
the out-of-plane momentum is $q_z=\pi$.}
\label{qw-map-D-pi}
\end{figure}

As we have discussed in Sec.~III C, each bond-charge susceptibility may also be defined by the 
projection of $D_{ab}$, not $D_{ab}^{-1}$, onto the corresponding eigenvectors 
$V^{d{\rm bond}}$, $V^{s{\rm bond}}$, and $V^{d{\rm CDW}}$. That is, 
$\tilde\chi_{d{\rm bond}} (\vq,\omega)=N(\delta/2)^{2}(D_{33}+D_{44}-2 D_{34})/2$, 
$\tilde\chi_{s{\rm bond}}(\vq,\omega)=N(\delta/2)^{2}(D_{33}+D_{44}+2 D_{34})/2$, 
and $\tilde\chi_{d{\rm CDW}}(\vq,\omega)=N(\delta/2)^{2}(D_{55}+D_{66} -2 D_{56})/2$. 
Their spectra, namely the imaginary part of each susceptibility, are shown in \fig{qw-map-D-pi}. 
The continuum spectrum is essentially the same as that shown in \fig{qw-map-D-1-pi} in Sec.~III C. 
A crucial difference is the presence of collective excitations from the on-site 
charge fluctuations, which remain strong along $(0,0)$-$(\pi,0)$ direction especially 
for  $\tilde\chi_{d{\rm CDW}}$. Because of symmetry there is no collective excitations 
along $(0,0)$-$(\pi,\pi)$ direction for $\tilde\chi_{d{\rm CDW}}$ and $\tilde{\chi}_{d{\rm bond}}$. 
Along $(\pi,0)$-$(\pi,\pi)$ direction the energy of the collective excitation becomes  
larger than $1 (=t)$. 
The presence of these collective excitations may be apparent from \fig{qw-map-allD-pi} 
where all elements of $D_{ab}$, 
in principle, contain the collective on-site charge excitations above the continuum spectrum 
as we have discussed in Sec.~III~A. Thus $\tilde{\chi}_{d{\rm bond}}$, $\tilde{\chi}_{s{\rm bond}}$, 
and $\tilde{\chi}_{d{\rm CDW}}$  necessarily have such contamination. 

\newpage
\bibliography{main} 
\end{document}